\documentclass[aps,amsmath,amssymb,arxiv,preprint]{revtex4-1}
\usepackage{amsmath, latexsym, color, graphicx, tabularx, booktabs, enumerate,upgreek}
\usepackage{multirow}
\usepackage[normalem]{ulem} 
\usepackage{caption}
\usepackage{epstopdf}
\usepackage{mathtools}
\usepackage{hyperref}
\usepackage{cleveref}
\begin{document}
\title{ Bending as a control knob for the electronic and optical properties of phosphorene nanoribbons}
\author{Bimal Neupane}
\email{bimal.neupane@temple.edu}
\author{Hong Tang }
\author{Niraj K. Nepal}
\author{Adrienn Ruzsinszky}

\affiliation{Department of Physics, Temple University, Philadelphia, Pennsylvania 19122, United States}

\maketitle

\section{Abstract}
We have assessed mechanical bending as a powerful controlling tool for the electronic structure and optical properties of phosphorene nanoribbons.  We use  state-of-the-art  density functional  approximations in our work. The overall performance of the recently developed  meta-generalized gradient approximation (meta-GGA) mTASK [Phys. Rev. Mater. \textbf{5}, 063803 (2021)] functional establishes the method as a useful alternative to the screened hybrid HSE06 for better band gaps  of phosphorene nanoribbons.  We  present a detailed  and novel analysis  to interpret the optical absorption of bent phosphorene nanoribbons using the GW-Bethe-Salpeter  approximation (GW-BSE).  We demonstrate the important role of the unoccupied in-gap state, and conclude that this in-gap state in armchair nanoribbons introduced by bending can significantly affect the properties of low-energy excitons, and add some useful opportunities for applications as optoelectronic device.

\section{Introduction}

 Phosphorene monolayer has attracted attention due to its remarkable physical and chemical properties. \cite{yang2015optical,pei2016producing,liu2014phosphorene,zhang2014extraordinary}. The electronic and optical properties of phosphorene can be modulated by defect \cite{wang2015native,kou2015anisotropic}, strain engineering \cite{rodin2014strain,quereda2016strong}, electric field\cite{liu2015switching,gao2017vastly} , edge functionalization \cite{peng2014edge,xie2017tuning}, and adatom adsorption\cite{hu2015first,ding2015structural}.
Tensile strain in a single layer (1L) phosphorene is fascinating, affecting bandgap narrowing, anisotropic phonon response, topological phase transition, \cite{xie2018electronic,wang2015remarkable,sisakht2016strain} etc. Watts et al. have  synthesized phosphorene nanoribbon (PNR) deposited on graphite by using ionic
scissoring of macroscopic black phosphorus crystals\cite{watts2019production}.
 The electronic properties of nanoribbons can be tuned with compressive and tensile strains. A tunable bandgap of phosphorene with bending is favorable for allowing flexibility in device design and applications. Recent studies with the Perdew-Burke-Ernzerhof (PBE) \cite{perdew1996generalized}  functional suggest that mechanical bending can be used to control the conductivity of phosphorene nanoribbons \cite{pandey2021flexoelectricity,yu2016bending}. Bending induces highly non-uniform local strain, resulting in changing electronic properties such as density of states, band structure, local charge density, etc\cite{pandey2021flexoelectricity,yu2016bending,nepal2019first}, and in addition resulting in unique phenomena not observed by homogeneous  linear strain.

Accurate prediction of fundamental band gaps from  semilocal density functional approximations for any material, especially for low-dimensional materials is one of the critical challenges in density functional theory (DFT)\cite{perdew1983physical}.  The mTASK meta-GGA\cite{neupane2021opening} is  based on the original form of the TASK\cite{aschebrock2019ultranonlocality} meta-GGA,   with enhanced  nonlocality so that mTASK has the screening appropriate for low-dimensional materials, competing with the accuracy of the more expensive screened hybrid HSE06 method \cite{neupane2021opening,tran2021band}.   Quasiparticle excitations are reliably captured by the GW approximation \cite{hedin1965new,hybertsen1985first}, while neutral excitations and the corresponding optical response are well described by GW plus Bethe-Salpeter equation (GW-BSE) \cite{rohlfing1998electron, albrecht1998ab,benedict1998optical}.  Excitonic effects are largely strengthened in low-dimensional materials with reduced screening.  \cite{choi2015linear,qiu2013optical}.

Bending was found to induce unoccupied in-gap states in armchair phosphorene nanoribbon  (APNR)\cite{yu2016bending}. A recent work on bending utilizing the GW-BSE approximation for armchair phosphorene suggests that the in-gap (edge) state may affect the behavior of low-energy excitons under strong inhomogeneous strain fields \cite{sun2020excitons}.    The bending scheme utilized in that work mainly focuses on compressive-only or tensile-only strains in nanoribbons. Such bent nanoribbons have to be conformed on a supporting media in order to maintain the bending configurations.   We use the bending scheme similar to the bending of a thin plate as used in  Ref. \cite{yu2016bending,nepal2019first,neupane2021opening,tang2021tunable}.  In our bending scheme, which is different from that of Ref.\cite{sun2020excitons}, the horizontal distance between the two outmost phosphorus atoms along the width direction is fixed and all other atoms are free to relax. The bent structure is free of any supporting media except for the clamping of the two edges. Such fixed width of nanoribbon includes strong inhomogeneous strain fields at large curvature because the inner and outer layers of bent phosphorene have compressive and tensile strains \cite{yu2016bending,nepal2019first,neupane2021opening,tang2021tunable}, respectively.  In contrast, Sun et al. \cite{sun2020excitons} have revealed the excitonic funnel effect in PNRs by tensile, and strong compressive strains only. In addition Ref [34], did not consider  the sepration of edge state from bulk bands at large curvatures\cite{yu2016bending}. In our  work, we mainly focus on how the band structures and the optical absorption vary in PNR for a wide range of bending curvatures.  We find that the band gap of armchair PNR shows a nonmonotonic changing trend with bending curvature, while the trend is monotonic for zigzag PNR. We report that the edge state of armchair PNR (APNR) significantly affects the behavior of low-energy excitons at large curvatures. We demonstrate that the absorption peaks of APNR generally shift toward the high energy direction (blue shift). In contrast,   ZPNR's show that the absorption peaks shift toward the low energy region (red shift) under mechanical bending.  In addition, the tunable behavior of the energy splitting of the singlet and triplet excitons with bending in the phosphorene nanoribbons is also revealed.

\section{Computational Details}
The PBE functional was used to optimize (or relax) the nanoribbon structures, and we utilized relaxed PBE structures to calculate the band structures, fixing ionic positions, using various   density functional approximations.
These calculations were performed in the Vienna \textit{ab initio} simulations package (VASP) \cite{kresse1996efficient,kresse1999ultrasoft}. The valence electrons of all elements are treated by the projector augmented wave (PAW) pseudo-potential method \cite{blochl1994projector}, which is recommended in the VASP manual.  The plane-wave energy cut-off is set as 500 eV for all calculations, leading to converged results. The Brillouin zone is sampled by a Gamma$-$centered mesh of 8$\times$1$\times$1 for nanoribbons. The self-consistent field (SCF) iterations for the total energy are converged within $10^{-5}$ eV.
A vacuum layer of more than 15 \AA \hspace{0.15cm} was inserted along  with both the width and thickness directions to prevent interactions between  the periodic images due to the long-range Coulomb interactions. The bent nanoribbons were constructed by fixing the distance between the two outermost edge atoms. One hydrogen atom is attached to every edge  atom in the nanoribbon. The edge atoms for the bent nanoribbons were only allowed to relax in the periodic direction. The lattice parameter along the periodic direction of the nanoribbon was also relaxed. All other atoms were fully relaxed until the force on each atom became less than 0.01 eV/\AA. For the optical properties, we have taken PBE structures from VASP and performed electronic relaxation to obtain the ground-state eigenstates and eigenvalues using  the Quantum Espresso (QE) \cite{giannozzi2009quantum} package.  The Berkeley GW code was used to perform  the G$_0$W$_0$+BSE calculations. For  Quantum Espresso (QE)+BerkeleyGW, we use a coarse kgrid of 1X1X8(1X1X10) and empty bands of 10 times more than the valence bands (for the G$_0$W$_0$ part) for APNR (ZPNR). We use 10  valence and 12 conduction bands, with dense k-point grids of 1X1X32 or 1X1X40 for APNR or ZPNR respectively (for the BSE part). We use a dielectric cutoff of 20 Ry and Gaussian smearing with a broadening constant of 50 meV in the optical absorption spectrum.  To  evaluate  the pure bending effects, we have neglected the electron-phonon coupling. We also  investigate the spin singlet-triplet splitting. Fig. 1 shows the flat and bent structures of PNRs. 

\begin{figure}[h!]
    \centering
    \includegraphics[scale=0.45]{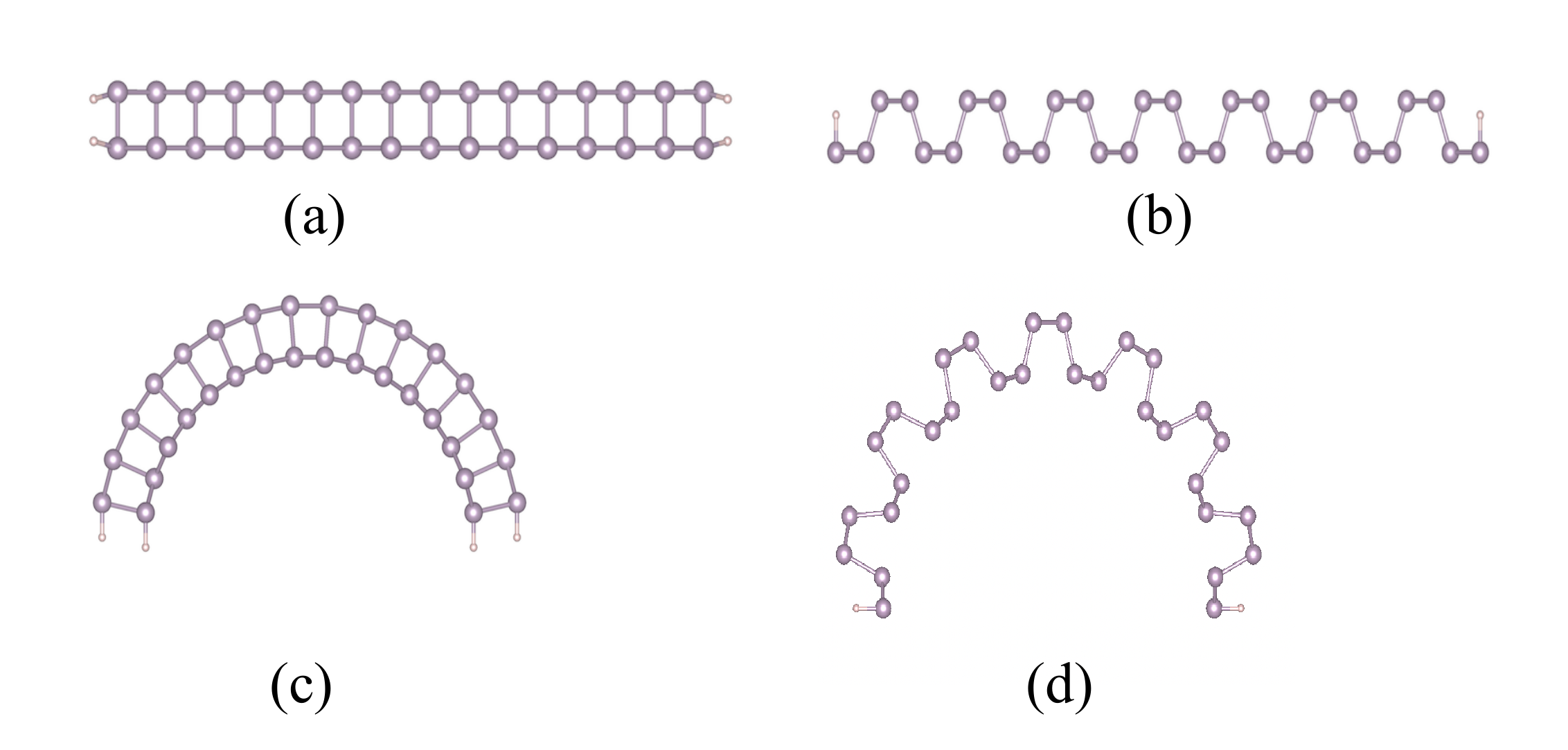}
    \caption{ Cross-sectional views of the (a) flat APNR, (b) flat ZPNR,  (c) bent APNR, and (d) bent ZPNR. }
\end{figure}

\section{Armchair phosphorene nanoribbons}
\subsection{Band gaps}

Fig. 2(a) shows the variation of band gaps along $\Gamma$ to X with respect to bending curvature using various DFT approximations. The PBE functional underestimates the band gaps of armchair phosphorene nanoribbons at different curvatures, as it has only the ingredient of density and the gradient of the density. The  strongly constrained and appropriately normed (SCAN)\cite{sun2015strongly} functional improves the band gaps compared to PBE by including some nonlocal exchange effects through the orbital-dependent ingredient $\alpha$.  Due to the increased nonlocality expressed by the derivative of exchange enhancement factor with $\alpha$, the TASK functional yields slight improvement compared to SCAN. The mTASK functional is designed to perform better for low-dimensional materials  by lifting the tight upper bound for one- or two-electron systems ($h_{X}(0)$ ) and by switching the limit of the interpolation function $(f_{X}(\alpha))$.  The band gaps of armchair nanoribbons from mTASK at different curvatures are improved and getting closer to G$_0$W$_0$@PBE.  However, the contribution of the edge states to the overall band structure from these semilocal and hybrid functionals slightly differs from the G$_0$W$_0$ results (See Fig. 2 and Fig. 4), which can affect the optical spectra especially for larger bending curvature. 
 
\begin{figure}[h!]
    \centering
    \includegraphics[scale=0.45]{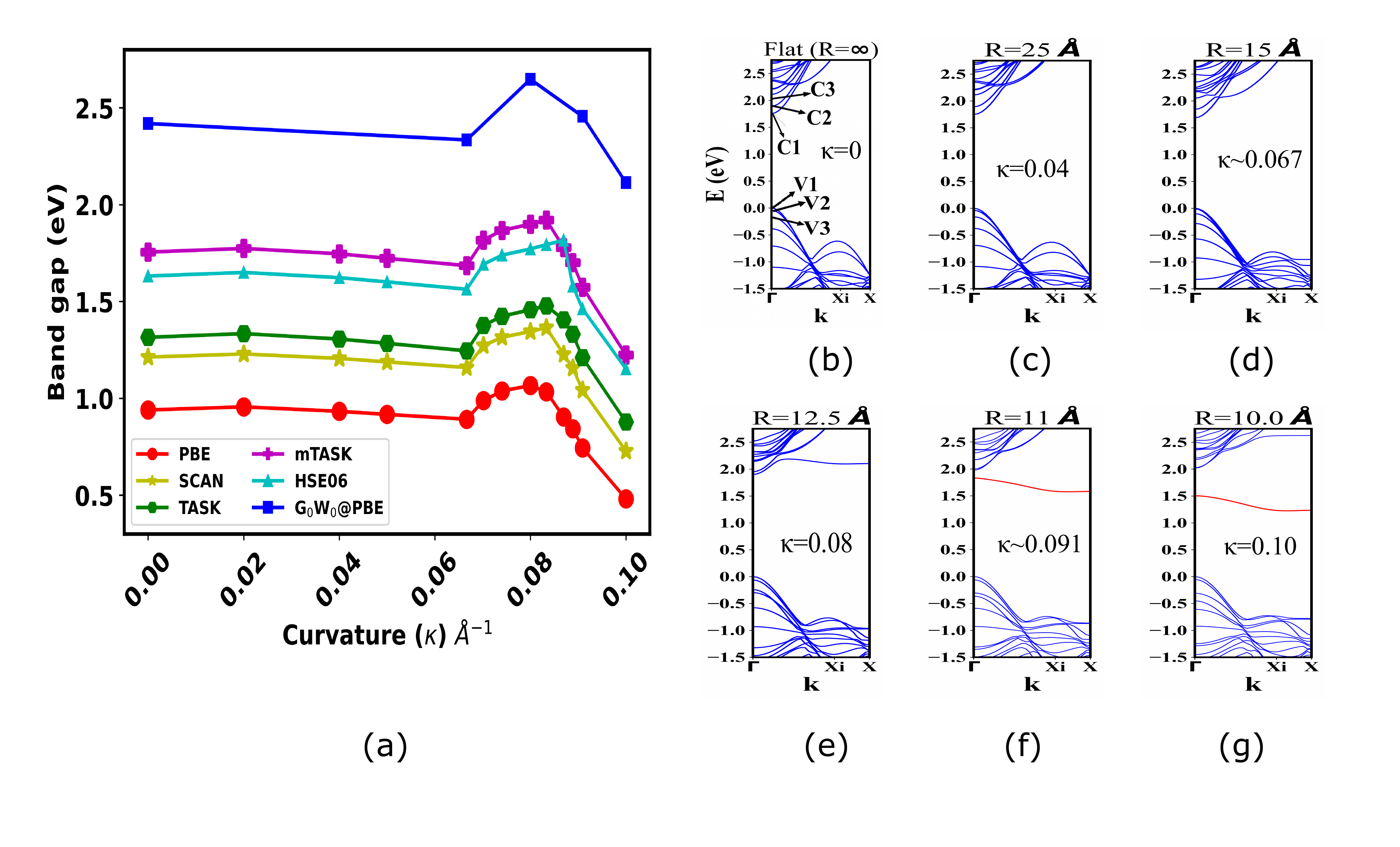}
    \caption{ (a) The band gap of armchair phosphorene nanoribbons at different bending curvatures with different density functional approximations. (b) to (g) represent   the band structure of armchair phosphorene nanoribbons at different radii of curvature (R) in \AA $~$ from the mTASK density functional.  Band energies are  are measured from the valence band maximum (VBM). The edge state is indicated by a solid red line. The eigenvalues on the y-axis are shifted such that the zero value represents the  (VBM). Bending curvature $\kappa = \frac{1}{R} (\AA^{-1}).$}
    \label{fig:band}
\end{figure}

\begin{figure}[h!]
    \centering
    \includegraphics[scale=0.40]{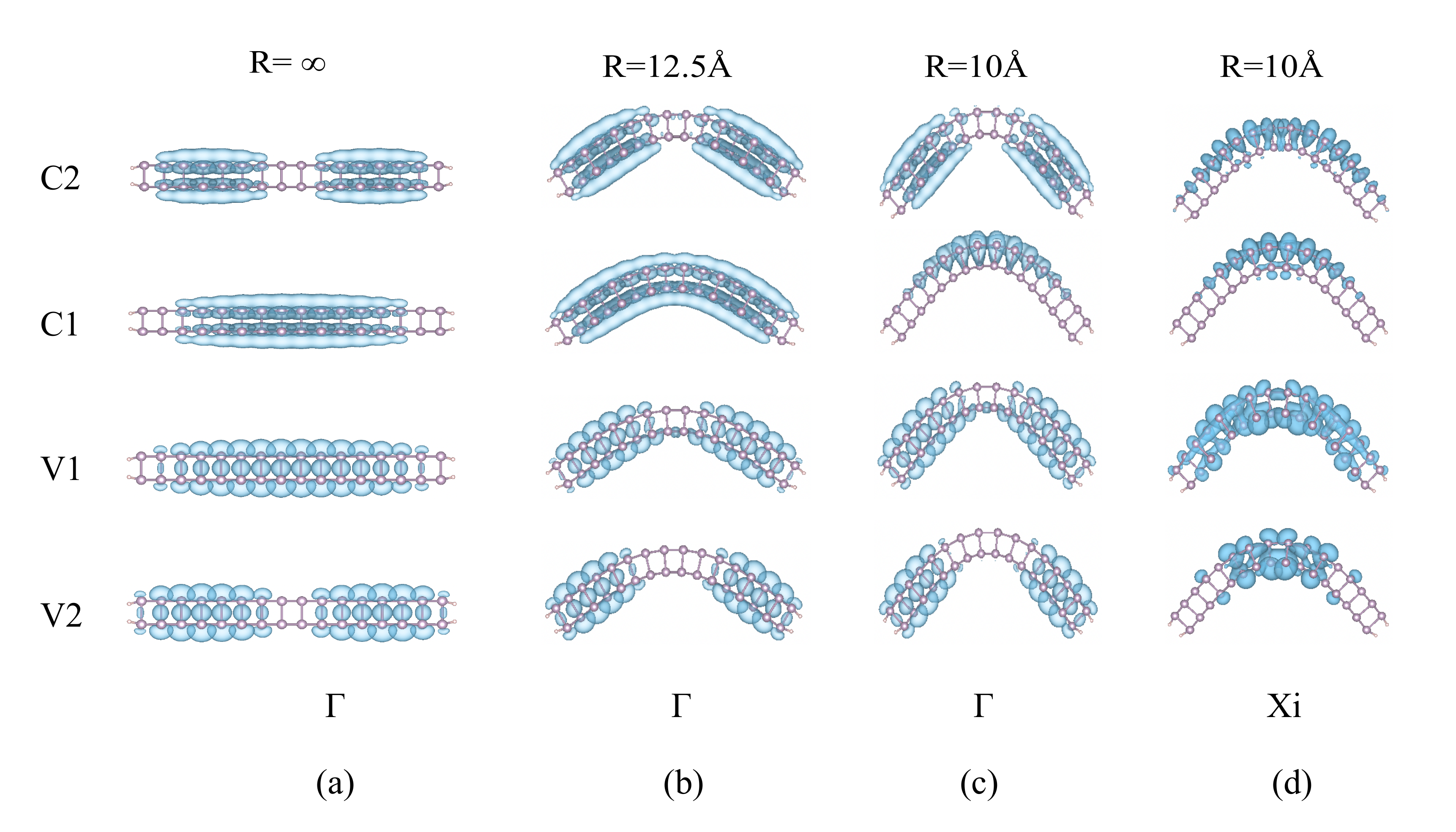}
    \caption{(a)-(c)Isosurfaces of partial charge densities of V1, V2, C1, and C2 states at  $\Gamma$ (0,0,0) at different radii of curvature (R) from the mTASK density functional  for APNR.  (d)  Isosurfaces of partial charge densities of V1, V2, C1, and C2 states at Xi (0.35,0,0) at radius of curvature (R =10 \AA) from the mTASK density functional.}
\end{figure}

\begin{figure}[h!]
    \centering
    \includegraphics[scale=0.45]{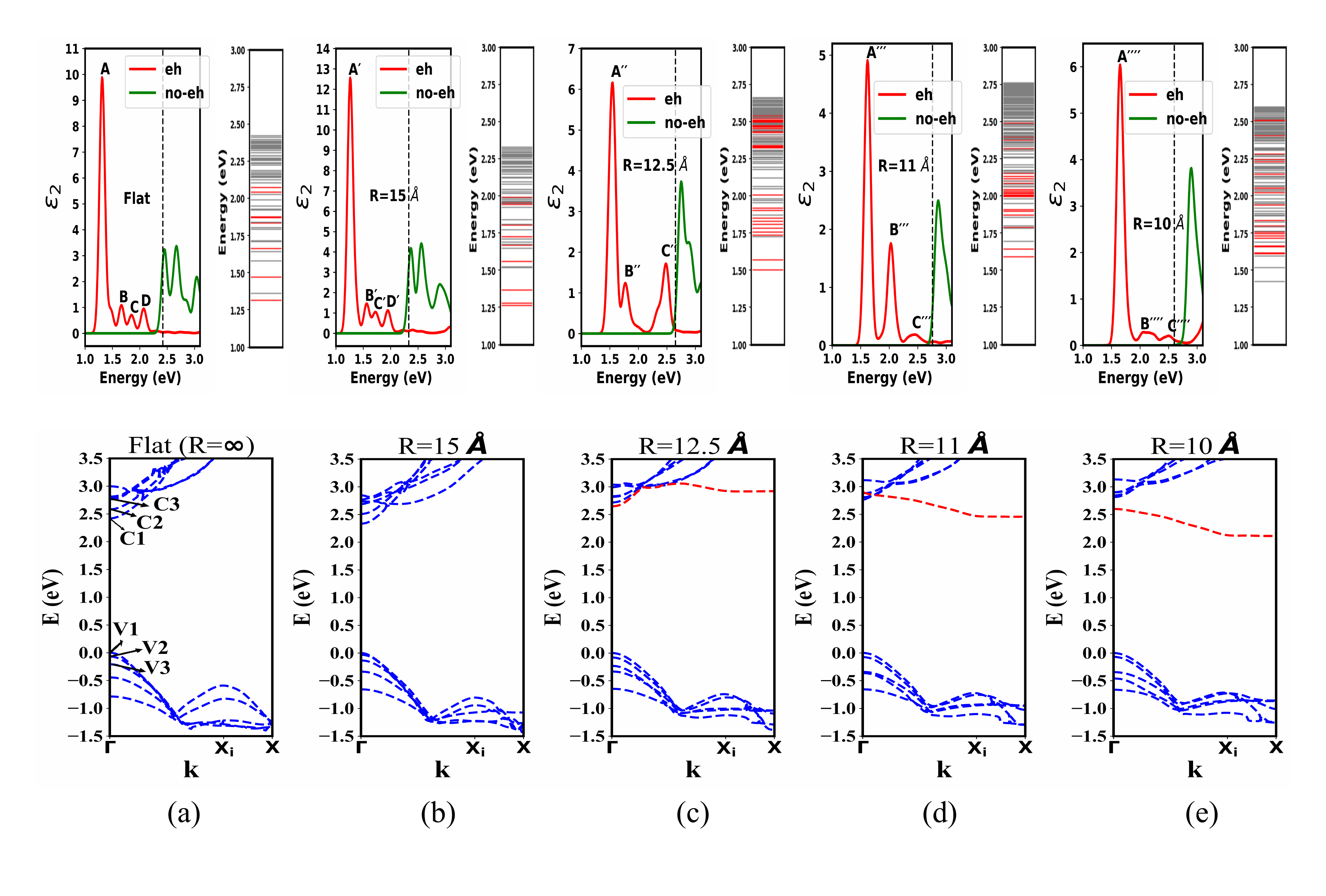}
    \caption{Absorption spectra and excitonic energy levels of armchair phosphorene at different radii of curvature (a) flat, (b) R = 15 \AA, (c) R = 12.5 \AA, (d) R = 11 \AA, and (e) R = 10 \AA (Upper panel).  The vertical black dashed line denotes the direct quasiparticle gap at $\Gamma$. Band energies are  measured from the VBM. Energy levels of optically bright exciton states are in red while those of dark states in gray. G$_0$W$_0$ band structures of APNR at different radii of curvature (Lower panel).}
    \label{fig:opti}
\end{figure}

Fig. 2(a) displays a slight decrease in the band gap from the flat APNR ( R = $\infty$ or $\kappa$=0) to curvature R = 15 \AA $~$ ($\kappa$ $\sim$0.067 \AA$^{-1}$). Further increase in curvature from R = 15 \AA $~$ $~$ ($\kappa$ $\sim$0.067 \AA$^{-1}$) to R $\sim$ 12.5 \AA $~$($\kappa$ $\sim$ 0.08 $\AA^{-1}$), results in band gap increase. The upshift of the conduction band minimum (CBM)  with bending from R = 15 \AA $~$to$~$ R $\sim$ 12.5 \AA $~$ is the main reason for the band gap increase. We define bands C1, C2, C3, V1, V2, V3, ... as in Figures~\ref{fig:band} and \ref{fig:opti}. From the flat nanoribbon to  R  $\sim$ 12.5 \AA, all functionals predict direct band gaps located at $\Gamma$. The upper edge state starts to separate from the continuous conduction bands at  R $\sim$ 12.5 \AA, forming indirect band gaps and leading to a unique behavior of APNR. This edge state is unoccupied and behave like acceptors. In the indirect band gap, the valence band maximum (VBM) is located at $\Gamma$ whereas the CBM is located at Xi (0.35,0,0). With further increase in the curvature from  R $\sim$ 12.5 ~\AA $~$ to R = 10 \AA, the indirect band gap decreases monotonically with curvature. The relaxed structure of nanoribbons breaks down with the curvature larger than R = 10 \AA $~$ \cite{yu2016bending}. Figs. 2(b) to 2(g) display the variation of band structure at the different radii of curvature from the mTASK functional.   Fig. 3 shows the partial charge densities of V1, V2, C1, and C2 states at $\Gamma-$point for different radii of curvature. Before the bending occurs, the charges on the top of the valence band (i.e., V1) are distributed over the entire ribbon width. As the bending curvature increases from infinity to R $\sim$ 12.5 \AA, the partial charges on the top of the valence band starts to localize at the edge and deplete in the middle ribbon region, giving the opportunity to alter the conductivity.  With further increase in the curvature from R = 12.5 \AA $~$ to R = 10 \AA, the partial charge on V1 is localized at the edge while the charges on conduction band (C1) become localized in the middle region of the nanoribbon at $\Gamma$. We have analyzed the charge distribution at Xi because C1 has the minimum value at this point. Both  C1 and V1 have the charge concentrated around the middle region of nanoribbon at Xi. In photocatalytic applications, such as water splitting, it is crucial that the shift of the edge states under bending is in favor of efficiency and preserves the photocatalytic properties of the semiconductors\cite{zhuang2013computational}.

\subsection{Optical spectra}
 
In order to reveal excitonic properties of bent APNR, the Bethe-Salpeter equation is solved on top of G$_0$W$_0$ calculations. Fig. 4 represents the absorption spectra, excitonic energy levels, and quasiparticle band gaps (G$_{0}$W$_{0}$@PBE) of bent APNR. For  the flat APNR, the first peak A consists of  a bright excitonic state of energy 1.31 eV, which is mainly formed by a transition between V1 to C1 and  V2 to C2 near $\Gamma$ point. The second peak B consists of exciton states that come from the transition V1 to C3 and V2 to C3 near $\Gamma$. The third peak C and the fourth peak D consist of bright excitons involving transitions of many valence and conduction bands that increase complexity for analysis.  The top and side views of modulus square of the exciton wavefunction corresponding to peak A are illustrated in Fig. 5(a). The exciton probability distribution of A spread over to the middle region. For R = 15 \AA $~$,   the main absorption peak  is slightly shifted toward the  low energy region (red shift), and the number of bright excitons  are increased. This red shift is related to  a slight decrease in the quasiparticle gap. Although the quasiparticle gap decreases compared to   the flat structure's gap, the increased number of bright excitons near A$^\prime$  play a crucial role in the overall increase  of the height of the absorption peak. The properties of the peaks at R = 15 \AA $~$ are similar to the ones of the flat nanoribbon. For R = 12.5 \AA,  which is a critical curvature, the edge state starts to appear, and the APNR is still a direct semiconductor. The direct quasiparticle gap at $\Gamma$ (2.65 eV, see Fig. 4(c)) is significantly lower than that of Xi (3.70 eV). We do not expect  a bright exciton forming from the transition of valence bands to conduction bands near Xi. Our detailed analysis  confirms that  the low-energy peaks come from exciton states  formed from the transition between valence bands to conduction bands near the $\Gamma$ point, not at Xi.  Fig. 5(b)  indicates that the exciton probability distribution of A$^{\prime\prime}$ is confined in the left region of the nanoribbon.

\begin{figure}[h!]
    \centering
    \includegraphics[scale=0.35]{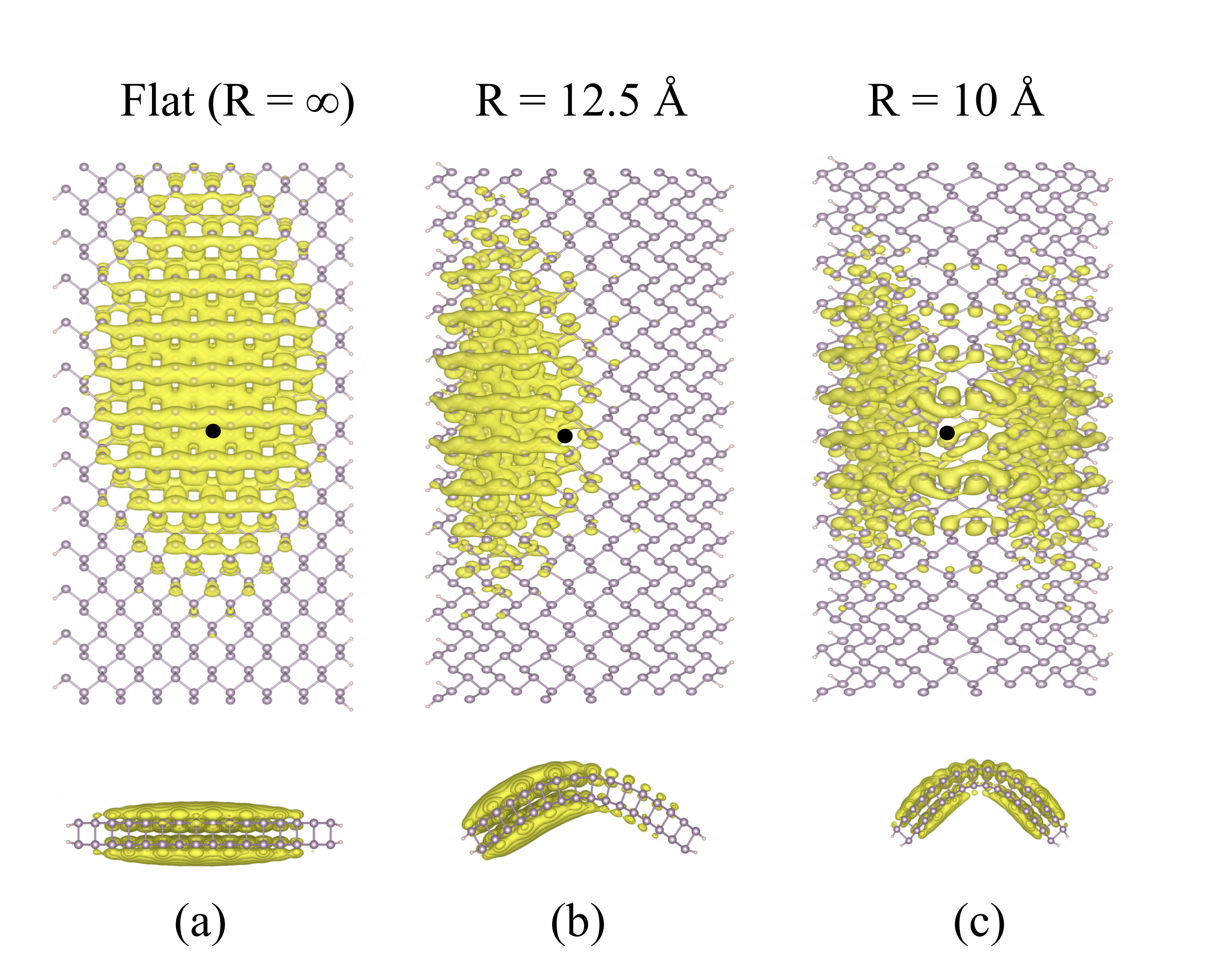}
    \caption{ Top view and side view of excitonic probability distribution function ($|\psi(\textbf{r}_e,\textbf{r}_h)|^2$) in real space with the hole fixed near the P atom (black dot). The bright excitonic states contributing to the first absorption peak are plotted. }
\end{figure}

For R=11 \AA, the quasiparticle band structure  is shown in Fig. 4(d), and confirms an indirect semiconductor due to the edge state. The first absorption peak A$^{\prime\prime\prime}$ consists of the combination of many bright exciton states at energy 1.67 eV, mainly transitions from V2 or V3 to C2 or C3 around $\Gamma$,   but the probability of transition from valence states to C1 around $\Gamma$ to form peak A$^{\prime\prime\prime}$ is negligible. The C1 band and valence bands near Xi are flatter. Flatter bands are preferred to form enhanced excitonic effects. A previous assessment  suggests  that flat bands contribute to a sizeable joint density of states, strengthening the chance of forming electron-hole (e-h) pairs \cite{cohen2016fundamentals}. The second peak B$^{\prime\prime\prime}$ consists of many bright exciton states at energy 2.03 eV. This peak appears from the transition from valence bands to the C1 edge state near Xi. This flat edge state can significantly tune the optical properties of phosphorene. After APNR becomes an indirect semiconductor, the contribution from the edge state to form bright excitons appears. 

\begin{table}[h!]
\caption { Quasiparticle band gaps G$_0$W$_0$$@\text{PBE}$ at $\Gamma$, the lowest absorption energy peak $E_{AI}$, and the excitonic binding energy for the lowest absorption peak ($E_{b}$= G$_0$W$_0$$@\text{PBE}$ - $E_{AI}$) of armchair phosphorene nanoribbons at different radii of curvature (R). The values of G$_0$W$_0$$@\text{PBE}$ in the parenthesis with $^{*}$ and $^{**}$ represent the direct gap from V2 to C1 at $\Gamma$ and the direct gap from V1 to C1 at Xi, respectively. }
\begin{tabular}{cccc}
\hline
\textbf{R (\AA)}   & \textbf{G$_{0}$W$_{0}$$@$PBE (eV)}   & \textbf{E$_{AI}$ (eV)}   &  \textbf{E$_{b}$ (eV)}  \\
\hline
\hline
$\infty$  &2.42	& 1.31 &	1.11  \\
15	& 2.33	& 1.26 &	1.07 \\
12.5 &	2.65 &	1.55 &	1.10\\
11 &	2.76 (2.81$^{*}$) (3.12$^{**}$) &	1.67 &	1.09 \\
10 &	2.60 (2.81$^{*}$) (2.85$^{**}$) &	1.65 &	0.95 \\
\hline
\end{tabular}
\end{table}

 The optical spectrum at R = 10 \AA $~$ emphasizes the role of edge state. At R =10 \AA, the first absorption peak A$^{\prime\prime\prime\prime}$ consists of four exciton states at energy 1.65 eV, mainly due to combinations of  transitions  from V1 to C1, from V2 to C1, and  from V3 to C1, around Xi,  from  V2 to C2 and from V1 to C3 near $\Gamma$  (see Supplementary Table S IV). C1 is the edge state which is entirely separated from continuum bulk states. For R = 11 \AA, only the second peak B$^{\prime\prime\prime}$ is formed from the transition of valence bands to  the edge band near Xi, whereas for R =10 \AA, both the first peak A$^{\prime\prime\prime\prime}$ and the second peak B$^{\prime\prime\prime\prime}$  have sizable contributions from the transition of valence bands to  the edge band near Xi.   Peak B$^{\prime\prime\prime\prime}$  also has contributions from the transition  from V1 to C6 near $\Gamma$, besides the transition   from the valence bands to C1 near Xi. Our results  show that, with increasing curvature  up to R =  10 \AA, the transitions  from the valence bands to C1 around Xi   appear both in the first and second highest peaks.  While increasing the curvature, the excitons formed from the transition  from valence bands to edge band near Xi extend to the lower energy region,  demonstrating the controlling power of bending on the optical response. Overall, the absorption peaks of APNR generally shift toward the high-energy region while increasing the curvature. With increasing curvature, the number of bright excitons created by the transition from valence bands to edge band increase.  Bright excitons can be formed by the transition from valence to edge band even when the quasiparticle gap becomes indirect,  corresponding to
the exciton wave function of A$^{\prime\prime\prime\prime}$ .  Peak A$^{\prime\prime\prime\prime}$ appears from a combination of four bright excitonic states (see Supplementary Fig. S5 for the separate plots of excitonic probability distributions of those excitonic states). The evolution of the exciton probability distributions in Fig. 5 with increasing bending supports the unique optical response of the transition from a direct to indirect semiconductor. The exciton wave function plots show that the excitons  are distributed  in the middle or edge region of the nanoribbon, depending on whether the contribution comes from valence to edge  bands near Xi or valence to bulk conduction bands near $\Gamma$.  This feature is consistent with the partial charge distribution of APNR at R=10 \AA $~$ at $\Gamma$ and Xi (Figs. 3(c) and 3(d)).
The calculated quasiparticle gaps, optical gaps, and exciton binding energies at different curvatures for the lowest exciton states are presented in Table I. 

\section{Zigzag phosphorene nanoribbons}

\subsection{Band gaps}

\begin{figure}[h!]
    \centering
    \includegraphics[scale=0.4]{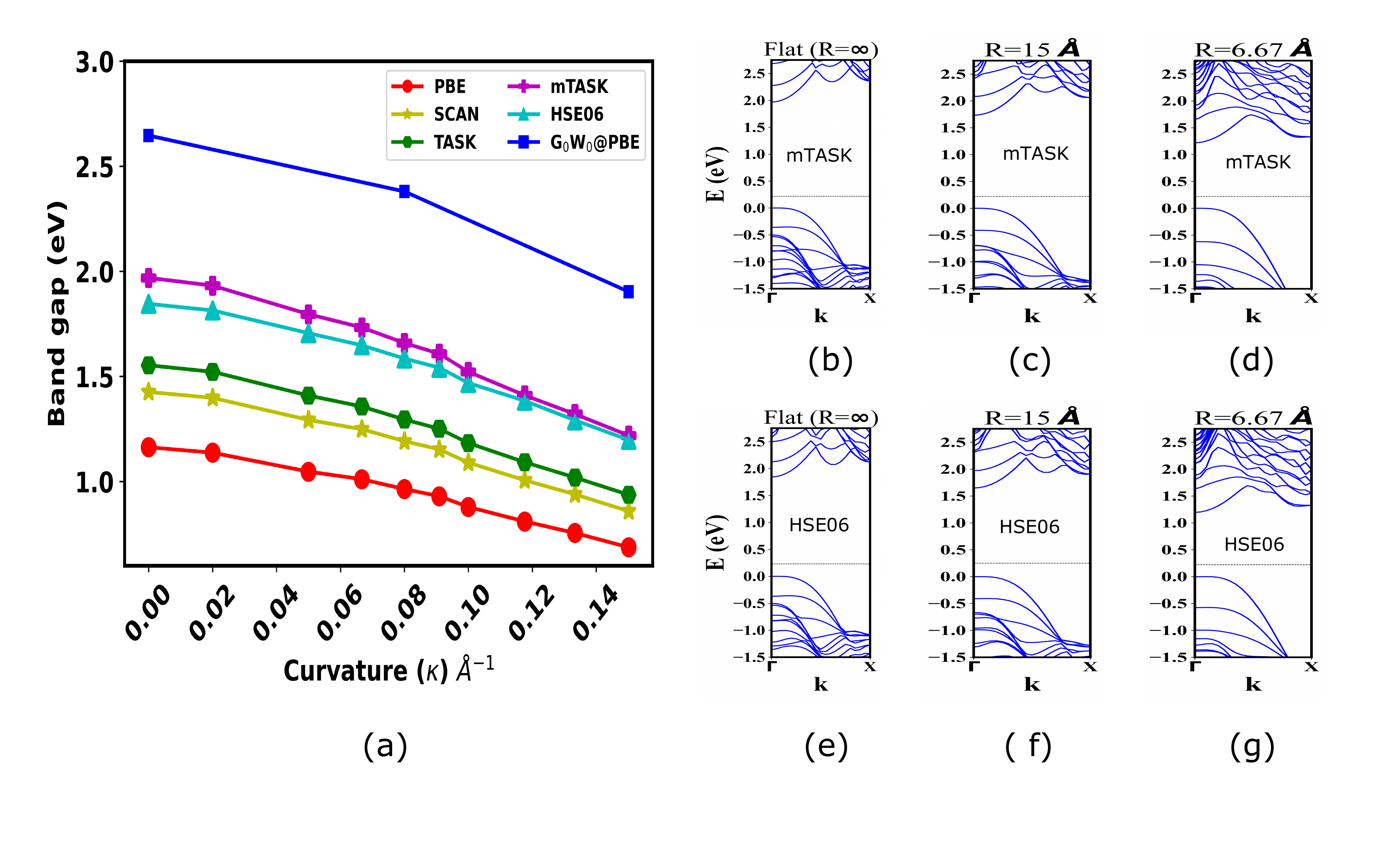}
    \caption{ (a) The band gap of zigzag phosphorene nanoribbons at different bending curvatures with different density functional approximations. Band structures of zigzag nanoribbon obtained from the mTASK density functional at different radii of curvature (b) flat R = $\infty$ (c) R = 15 \AA\ (d) R = 6.67 \AA\ and the hybrid HSE06 functional at different radii of curvature (e) flat R = $\infty$ (f) R = 15 \AA\ (g) R = 6.67 \AA\ . The Fermi-level is indicated by the dashed black line. Bending curvature $\kappa = \frac{1}{R}.$}
\end{figure}

Fig. 6(a) shows the variation of band gaps with  curvature from different functionals.   mTASK remains a better choice than TASK for band gap prediction, as has been found for APNR, and in general for materials with p bands \cite{neupane2021opening,tran2021band}.  Fig. 6, demonstrates  that mTASK is a successful alternative to HSE06, as it improves the latter method, bringing the band gaps closer to the ones of G$_{0}$W$_0$@PBE. As the bending curvature increases, the difference in band gaps from mTASK and HSE06 decreases. At large curvature, the mTASK and HSE06 functional predict almost the same band gaps. The shifting of the conduction bands at $\Gamma$ toward the Fermi-level for the mTASK functional is faster than the one of HSE06 functional,  explaining this trend as shown in Figs. 6(b) to 6(g).  Although the band gaps significantly changes with the bending curvature, all functionals considered in our work qualitatively predict a similar band pattern.

\subsection{Optical spectra}

\begin{figure}[h!]
    \centering
    \includegraphics[scale=0.75]{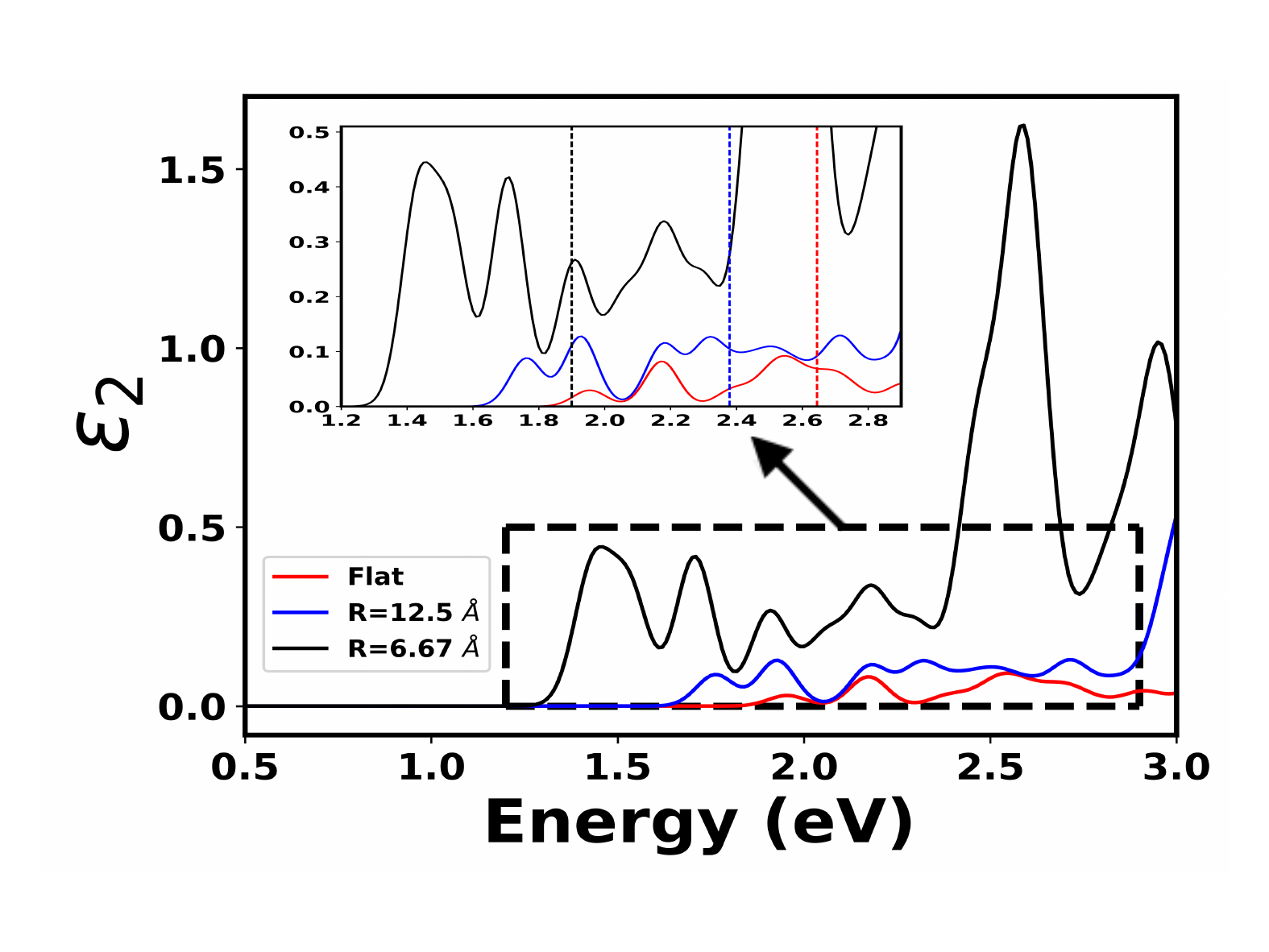}
    \caption{Absorption spectra of zigzag phosphorene at different radii of curvature with electron-hole interactions. The vertical dashed red, blue, and black lines denote the quasiparticle gap\sout{s} of zigzag  phosphorene nanoribbons for  the flat nanoribbon, for R=12.5 \AA, and R=6.67 \AA, ~ respectively.}
\end{figure}

The absorption spectra of zigzag phosphorene ribbons with electron-hole (e-h) interactions are shown in  Fig. 7. For a flat zigzag phosphorene ribbon,  a weak absorption occurs at photon energies 1.96 eV, 2.17 eV, and 2.55 eV. While increasing the  curvature, the absorption peaks shift toward low-energy region (red shift). When R = 12.5 \AA, absorption occurs at photon energies 1.76 eV, 1.93 eV, and 2.55 eV.  With further increasing the curvature i.e., R = 6.67 \AA, the strong absorption occurs at photon energies 1.46 eV and 1.71 eV. The optical absorption shifts toward a low-energy direction with bending (red shift). The red shift can be attributed to the direct quasiparticle fundamental gap decrease with bending. We also notice that the  height of the absorption peak of the lowest bright excitonic state when R = 6.67 \AA $~$    is about four orders of magnitude larger than   the one of the flat nanoribbon. The excitonic binding energies of the lowest absorption excitonic peaks are 0.69, 0.62, and 0.44 eV for  the flat nanoribbon, for R = 12.5, and R = 6.67 \AA, respectively.   The inset of Fig. 7 shows that, the third absorption peak is the highest peak for  the flat phosphorene, the second excitonic peak is the highest peak for  the R = 12.5 \AA ~ phosphorene, and the first excitonic peak is the highest peak for R = 6.67 \AA. The basic shape of the optical absorption strongly changes with bending,  confirming the controlling power of bending for optoelectronic applications.

\section{ Exchange splitting in bent phosphorene nanoribbons}

 We aim to demonstrate that the optoelectronic response trend in bent APNR's transfers to exciton exchange splitting with a great opportunity to control emitting states. Large exchange splitting has been noticed in Si quantum dots and doped Si nanowires. A certain analogy can be noticed when comparing the latter systems with our bent APNR's. Large exchange splitting enhances luminescence and overall the optical gain. Impurity or surface dangling bond-introduced occupied or unoccupied states qualitatively resemble of bending induced edge states in APNR. Edge states are found more localized than the bulk bands in nanoribbons, as impurity states are as well. Exciton localization, however can be altered in APNR's by bending. Here we find an extreme sensitivity of the exchange splitting with respect to the curvature and associated band structure in APNR.  Fig. 8  demonstrates the exchange splitting and curvature relation. To match the singlet-triplet splitting  with our prior analysis of the optical response, we choose the same bending curvatures to evaluate the exchange splitting.
The corresponding singlet-triplet splitting exhibits an interesting pattern. Ref.\cite{palummo2010giant} reports an exchange splitting of ~ 120 meV as being very large. While we do not see these large values for the flat structure and for R = 15,  12.5 and 10  \AA's , we identify a large exchange splitting for the peak A$^{\prime\prime\prime}$ at R =  11 \AA. This large splitting originates in the emergence of the localized edge state at this particular bending curvature. With further bending, however, the exchange splitting decreases again.  The exchange splitting in APNR is larger than the one of ZPNR  (see  Supplementary Table S VI for ZPNR). The shorter width of APNR than ZPNR results in increased overlapping between  the electron-hole wave functions, leading to an increase of exchange interaction in the BSE kernel.  Furthermore, we believe that the anisotropy along the armchair and zigzag directions of the phosphorene nanoribbon plays a role in the difference of the singlet-triplet splitting in APNR and ZPNR.    Obviously, the singlet-triplet splitting exhibits a strong and very delicate tunability with bending as a potential tool to control luminescence.

\begin{figure}[h!]
    \centering
    \includegraphics[scale=0.30]{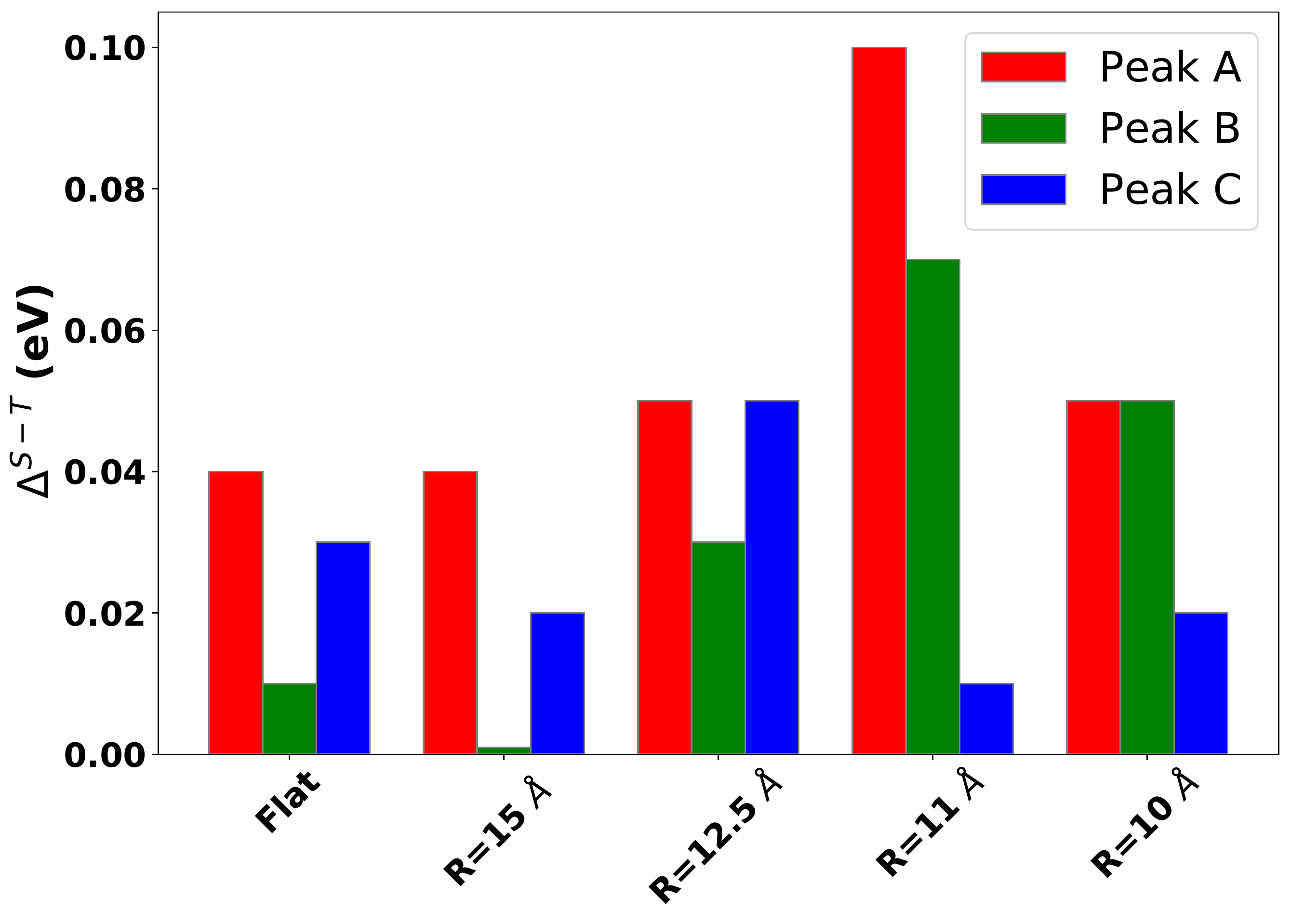}
    \caption{The variation of singlet-triplet splitting of the first three absorption peaks with curvatures for armchair nanoribbons. }
\end{figure}

\section{Conclusions}

In order to reveal the effect of bending, we have systematically investigated the electronic and optical properties of phosphorene nanoribbons from first-principles.
The mTASK meta-GGA approximation has been established for band gaps of low-dimensional materials recently. The DFT band gaps of PNRs obtained from the mTASK functional remain competitive with the screened hybrid HSE06, and get closer to G$_0$W$_{0}$@PBE the reference value. Our calculations show that (1) the band gaps and band dispersion of phosphorene nanoribbons strongly depend on bending curvatures, and (2) reveal the relevance of the edge states in the quasi-one-dimensional nanoribbons.

The GW-BSE calculations of armchair phosphorene at different curvatures reveal that the optical absorption can be significantly altered by bending, and the edge states play the role of a strong controlling tool for potential optoelectronic applications. Our work suggests that the lowest-energy exciton is bright or dark, depending on whether the edge state is entirely separate from the conduction band continuum or not. In addition, we reveal that the bright exciton can form from the direct transition from valence bands to the edge band when the APNR becomes an indirect semiconductor.  The brighter excitons come from the transition from valence bands to edge state with the increase of the curvature, and those excitons extend toward the low-energy directions. Our analysis suggests that the optical absorption peaks of ZPNR shift toward the low-energy region, and the height of the absorption peaks increase while increasing the bending curvature.
Furthermore, our work reveals the unique exchange splitting of APNR, with a maximum in the singlet-triplet energy differences as functions of the bending curvature. The large excitonic binding energy, the tunable optical gaps and exchange splitting of PNR make these materials promising candidates for optoelectronic device applications.

\section{Acknowledgment}
 This material is based upon work supported by the U.S. Department of Energy, Office of Science, Office of Basic Energy Sciences, under Award Number DE-SC0021263. The calculations were carried out on HPC resources supported in part by the National Science Foundation through major research instrumentation grant number 1625061 and by the US Army Research Laboratory under contract number W911NF-16-2-0189.

%\bibliography{lit}

\section{Supplementary Materials}
\subsection{Few-layer phosphorene}

For the few-layer phosphorene, AA stacking means that the top phosphorene layer is directly stacked on the bottom layer of phosphorene. AB stacking can be obtained by shifting either the top or bottom layer of phosphorene by half of the lattice constant along the a or b direction. Previous studies suggest that the AB stacking structure has the minimum energy. We consider AB stacked geometry. For the 2-, 3-, 4-, and 5-layer phosphorene, the stacking orders are AB, ABA, ABAB, and ABABA, respectively.

\begin{figure}[h!]
 \renewcommand\thefigure{S1}
    \includegraphics[scale=0.7]{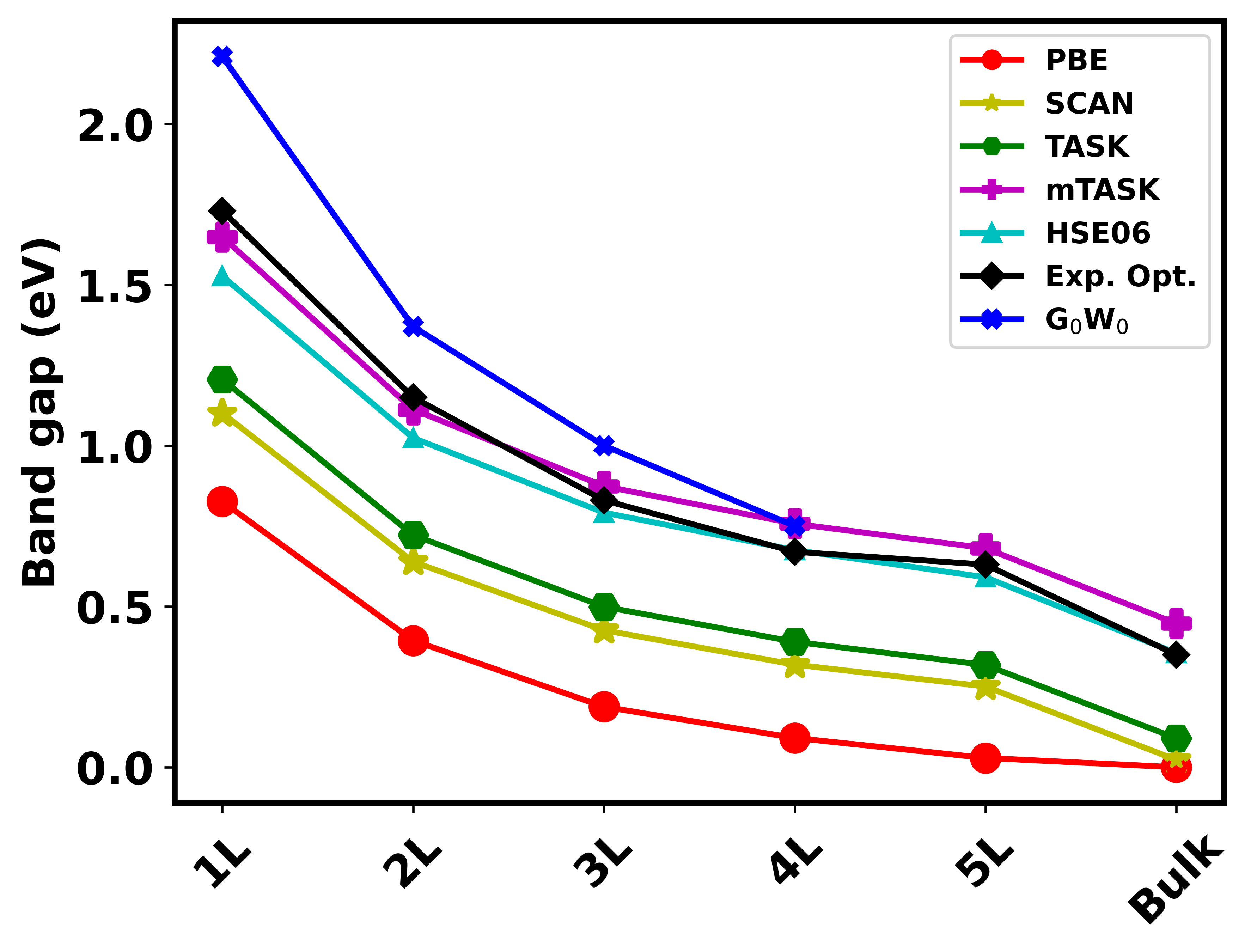}
    \caption{The band gap  of the few-layer phosphorene from different density functional approximations.}
\end{figure}

\begin{table}[h!]
 \renewcommand\thetable{S I}
\caption {Calculated band gaps of few-layer phosphorene and bulk phosphorous from different density functional approximations. The experimental optical gaps are also shown. $E_{b}$ represents the binding energy of the lowest energy exciton obtained from GW-BSE calculations. All values are in eV.}
\begin{tabular}{cccccccc}
\hline
\textbf{Number   of layers} & \textbf{PBE} & \textbf{SCAN} & \textbf{TASK} & \textbf{mTASK} & \textbf{HSE06} & \textbf{Exp. Optical Gap} &\textbf{G$_{0}$W$_{0}$}\\
\hline
\hline
\textbf{1L}                 & 0.83         & 1.10          & 1.21          & 1.65           & 1.53           & 1.73 \cite{ li2017direct}    & 2.21 \cite{ qiu2017environmental}             \\
\textbf{2L}                 & 0.39         & 0.64          & 0.72          & 1.11           & 1.02           & 1.15 \cite{ li2017direct}             &1.37 \cite{ qiu2017environmental}      \\
\textbf{3L}                 & 0.19         & 0.43          & 0.50          & 0.87           & 0.79           & 0.83  \cite{ li2017direct}               &1.00 \cite{ qiu2017environmental}   \\
\textbf{4L}                 & 0.09         & 0.32          & 0.39          & 0.76           & 0.68           & 0.67 \cite{zhang2017infrared}       &0.74 \cite{qiu2017environmental}         \\
\textbf{5L}                 & 0.03         & 0.25          & 0.32          & 0.68           & 0.59           & 0.63\cite{zhang2017infrared}      & --  \\
\textbf{Bulk}               & 0.00         & 0.02          & 0.09          & 0.45           & 0.36           & 0.35\cite{zhang2017infrared} &-- \\
\hline
\end{tabular}
\end{table}

 \begin{figure}[h!]
 \renewcommand\thefigure{S2}
    \includegraphics[scale=0.5]{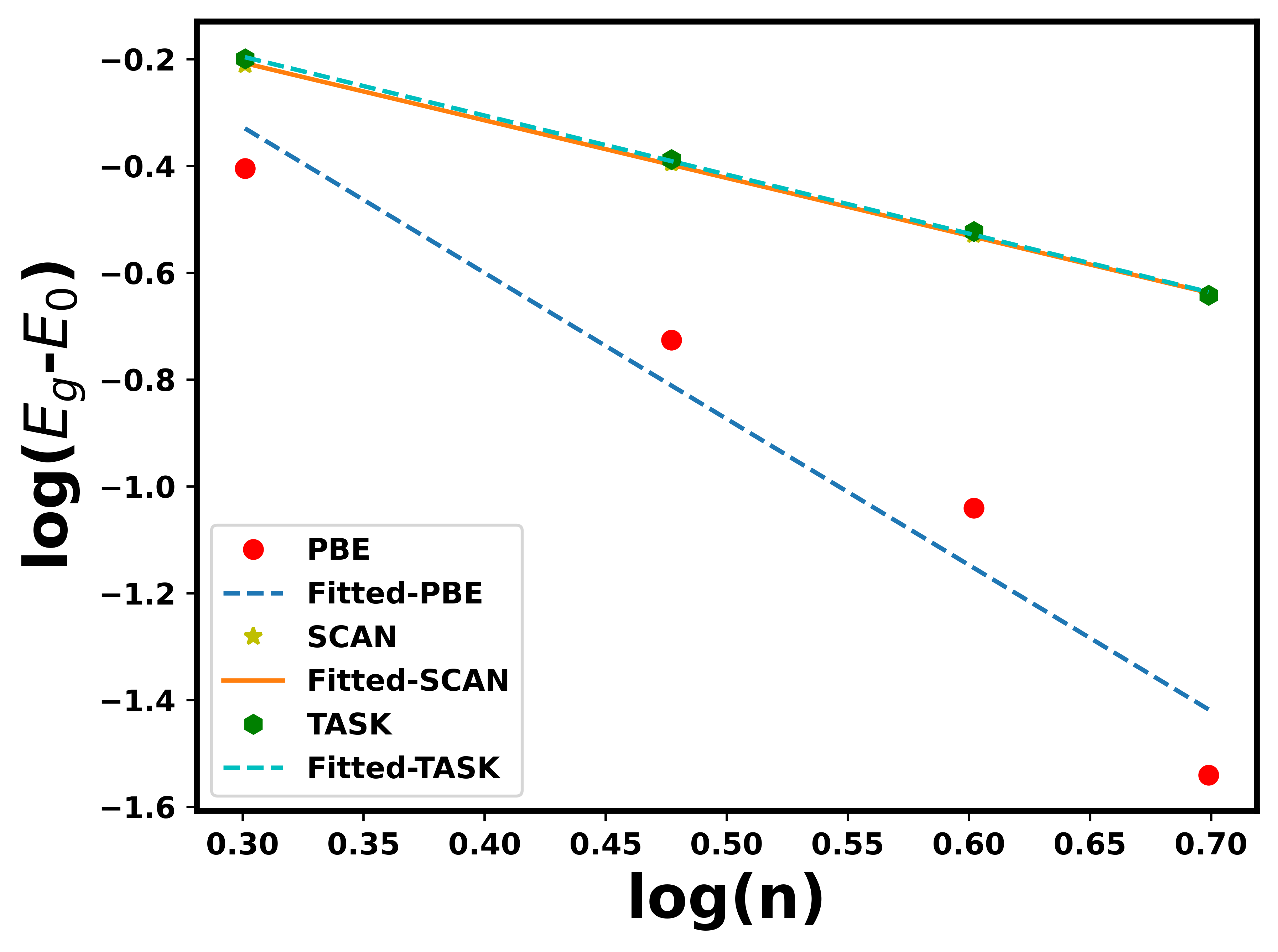}
    \includegraphics[scale=0.5]{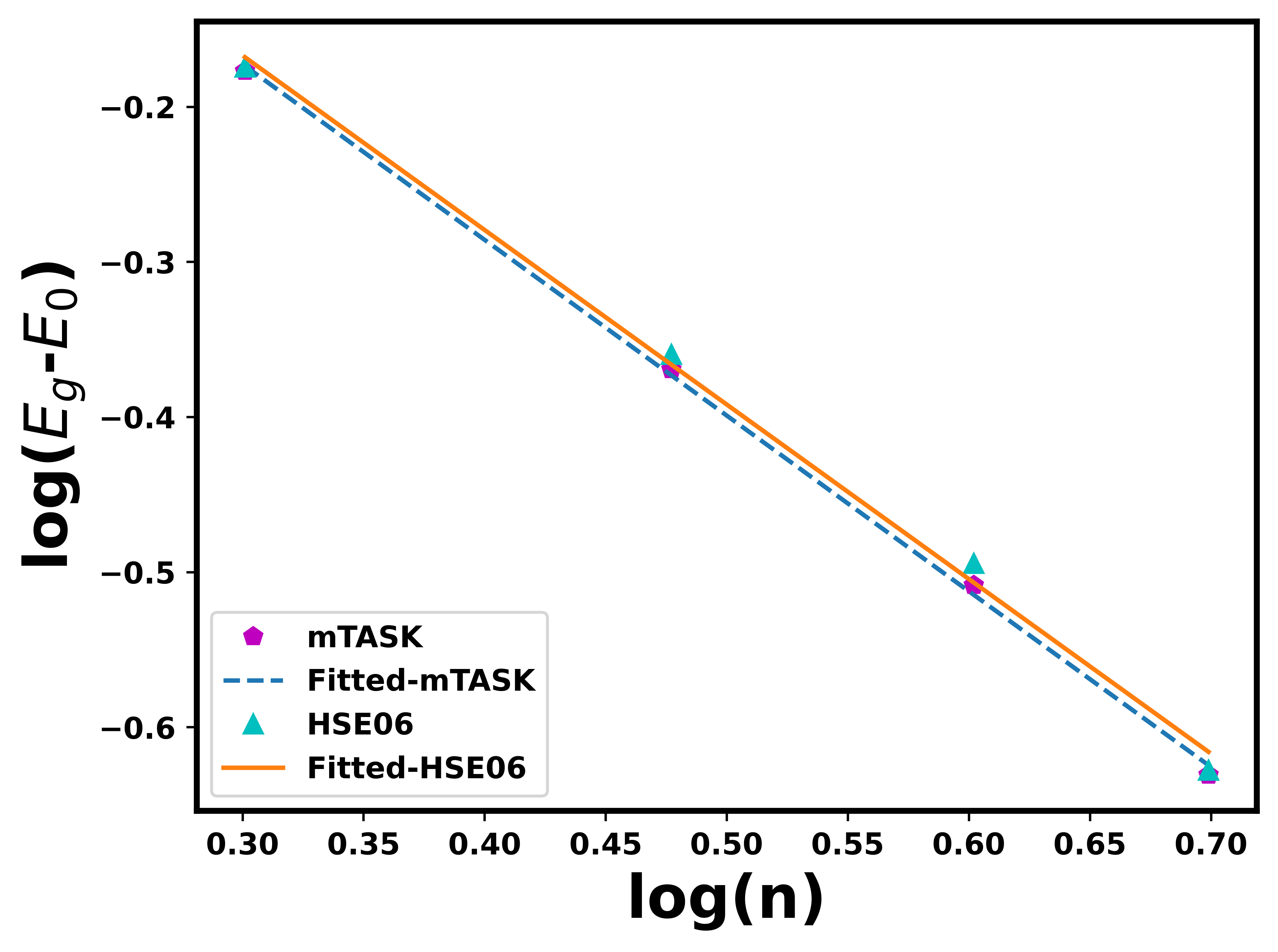}
  \caption{ The band gap ($E_{g} -  E_{o}$) in eV from different density functionals vs number of layers on a log-log scale.  }
% \label{fig:coffee}
\end{figure}

 The relation between the band gap of few-layer phosphorene and the number of layers is given by \cite{cai2014layer}:

\begin{equation}
    E_{g}=E_{0}+C/n^{\beta}   
\end{equation}
\begin{equation}
\Rightarrow log(E_{g}-E_{0})=log(C) -\beta~ log(n)
\end{equation}

where $E_{g}$ is the band gap of the few-layer phosphorene, $ E_{o}$ is the band gap of bulk phosphorene, and n is the number of layers in the few-layer phosphorene. As the number of layers decreases, the band gap increases following equation 1.

\begin{table}[h!]
 \renewcommand\thetable{S II}
\caption { The fitted parameters C and $\beta$ for band gaps of few-layer phosphorene from different density functional approximations according to the formula $log(E_{g}-E_{0})$=$log(C) -\beta~ log(n)$.}
\begin{tabular}{ccccc}

\hline
Functional & C     & $\beta$  &  &  \\
\hline
\hline
  PBE   & 0.321 & 2.735 &  &  \\
  SCAN       & 0.763 & 1.08  &  &  \\
  TASK       & 0.730  & 1.105 &  &  \\
   mTASK      & 0.680  & 1.133 &  &  \\
   HSE06      & 0.675 & 1.125 &  & \\
\hline
\end{tabular}
\end{table}

\begin{figure}[h!]
  \renewcommand\thefigure{S3}
    \includegraphics[scale=0.35]{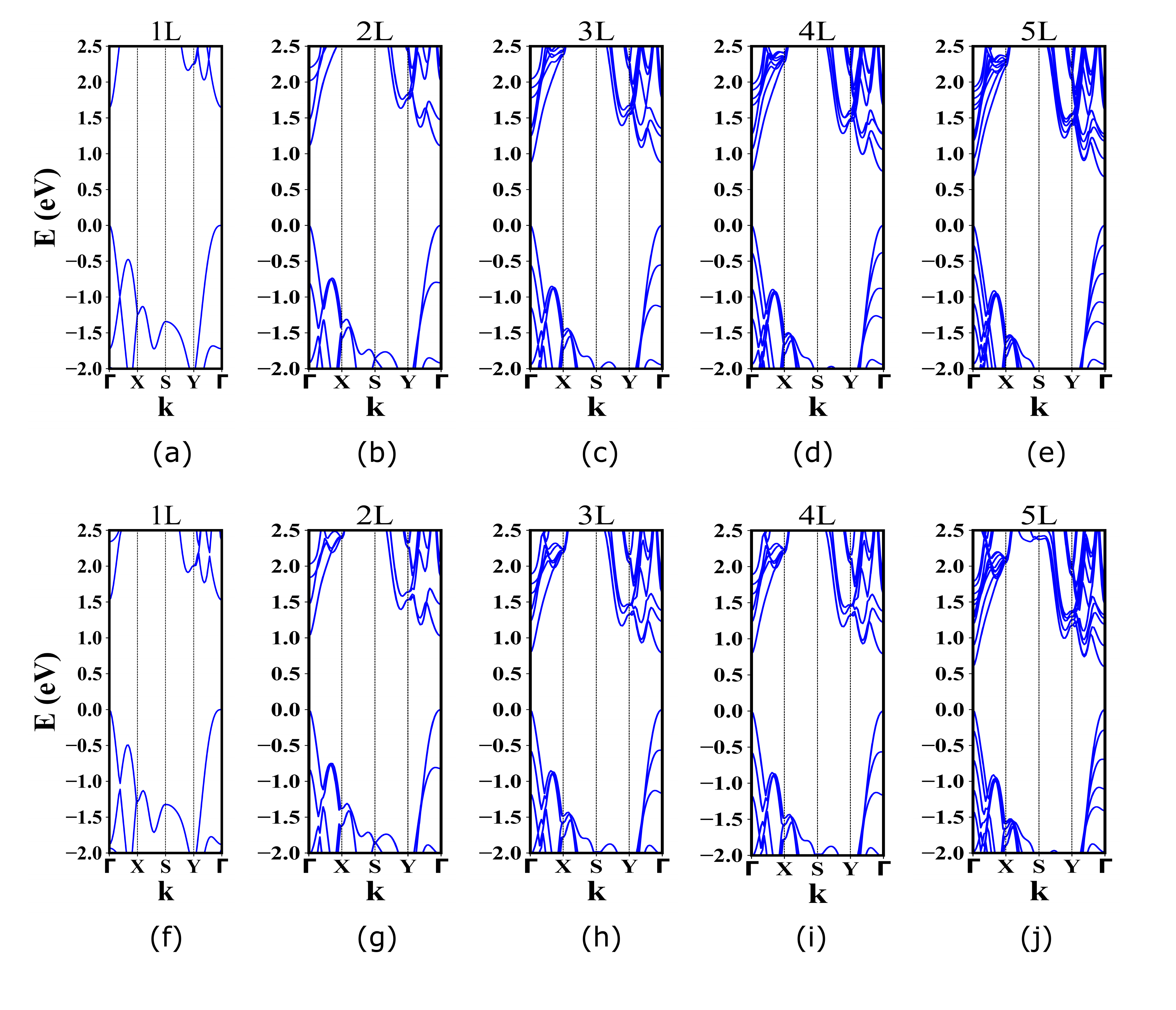}
    \caption{Band structures of few-layer phosphorenes obtained from the mTASK density functional (a) monolayer (1L) (b) bilayer (2L) (c) trilayer (3L) (d) tetralayer (4L) (e) pentalayer (5L) and the hybrid HSE06 functional (f) monolayer (1L) (g) bilayer (2L) (h) trilayer (3L) (i) tetralayer (4L) (j) pentalayer. }
\end{figure}

Figure S3 shows a comparison of the band structures of 1L to 5L phosphorene from the mTASK and the hybrid HSE06 functional. Both mTASK and HSE06 functionals predict direct band gaps at the $\Gamma$ point. The pattern of the band structure is similar to all few-layer phosphorenes. The asymmetric shape of the band dispersion around $\Gamma$ point suggests highly anisotropic properties of the electronic structure of few-layer phosphorenes. TASK and mTASK also suggest those anisotropic properties of the electronic structure of few-layer phosphorenes.

\begin{table}[h!]
 \renewcommand\thetable{S III}
\caption{Band gaps (eV) of armchair and zigzag phosphorene nanoribbons at different radii of curvature (R in \AA) from different density functional approximations.}
\begin{tabular}{ccccccc}
\hline
\hline
\textbf{Nanoribbons}            & \multicolumn{1}{c}{\textbf{R}} & \multicolumn{1}{c}{\textbf{PBE}} & \multicolumn{1}{c}{\textbf{SCAN}} & \multicolumn{1}{c}{\textbf{TASK}} & \multicolumn{1}{c}{\textbf{mTASK}} & \multicolumn{1}{c}{\textbf{HSE06}} \\
\hline
\hline
\multirow{14}{*}{\textbf{Armchair}}  &  $\infty$  & 0.94	&1.21	&1.32&	1.76&	1.63  \\
&50&	0.96&	1.23&	1.33&	1.77&	1.65\\
&25&	0.93&	1.21&	1.31&	1.75&	1.62\\
&20&	0.92&	1.19&	1.28&	1.72&	1.60\\
&15&	0.89&	1.16&	1.25&	1.69&	1.56\\
&14.25&	0.99&	1.27&	1.38&	1.82&	1.69\\
&13.5&	1.04&	1.32&	1.42&	1.87&	1.74\\
&12.5&	1.07&	1.35&	1.46&	1.90&	1.77\\
&12&	1.03&	1.37&	1.48&	1.92&	1.79\\
&11.5&	0.90&	1.23&	1.41&	1.78&	1.82\\
&11.25&	0.84&	1.16&	1.33&	1.70&	1.58\\
&11&	0.74&	1.04&	1.21&	1.57&	1.46\\
&10&	0.48&	0.73&	0.88&	1.22&	1.15\\
                                \hline
\multirow{10}{*}{\textbf{Zigzag}} & $\infty$ & 1.16&	1.43&	1.55&	1.97&	1.85 \\
&50&	1.14&	1.40&	1.52&	1.93&	1.81\\
&20&	1.05&	1.29&	1.41&	1.80&	1.71\\
&15&	1.01&	1.25&	1.36&	1.73&	1.65\\
&12.5&	0.96&	1.19&	1.30&	1.66&	1.59\\
&11&	0.93&	1.15&	1.25&	1.61&	1.54\\
&10&	0.88&	1.09&	1.18&	1.52&	1.47\\
&8.5&	0.81&	1.01&	1.09&	1.41&	1.38\\
&7.5&	0.76&	0.94&	1.02&	1.32&	1.29\\
&6.67&	0.69&	0.86&	0.94&	1.22&	1.20\\
                                \hline
\hline
\end{tabular}
\end{table}

\begin{figure}
  \renewcommand\thefigure{S4}
    \includegraphics[scale=0.50]{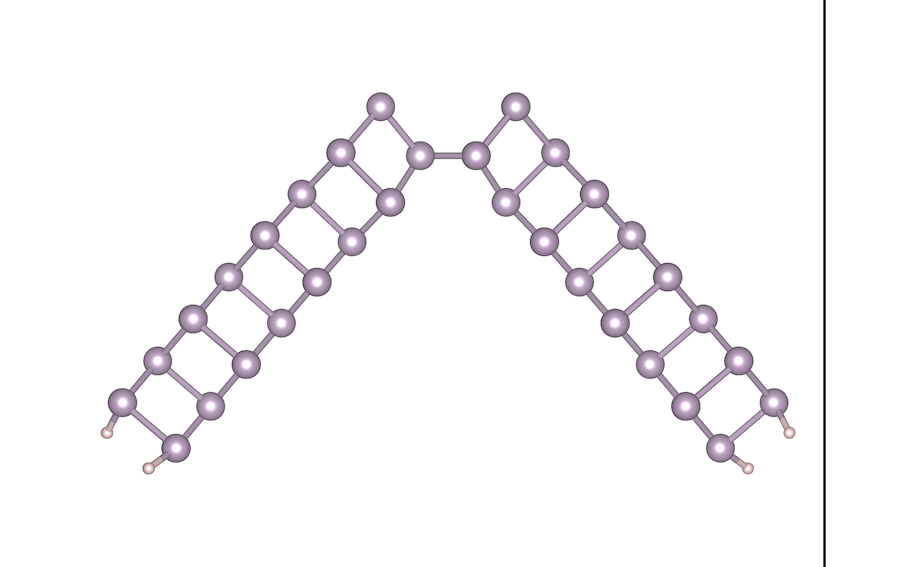}
    \caption{Broken structure of APNR after relaxation at very large curvature.}
\end{figure}

\begin{figure}[h!]
 \renewcommand\thefigure{S5}
    \includegraphics[scale=0.45]{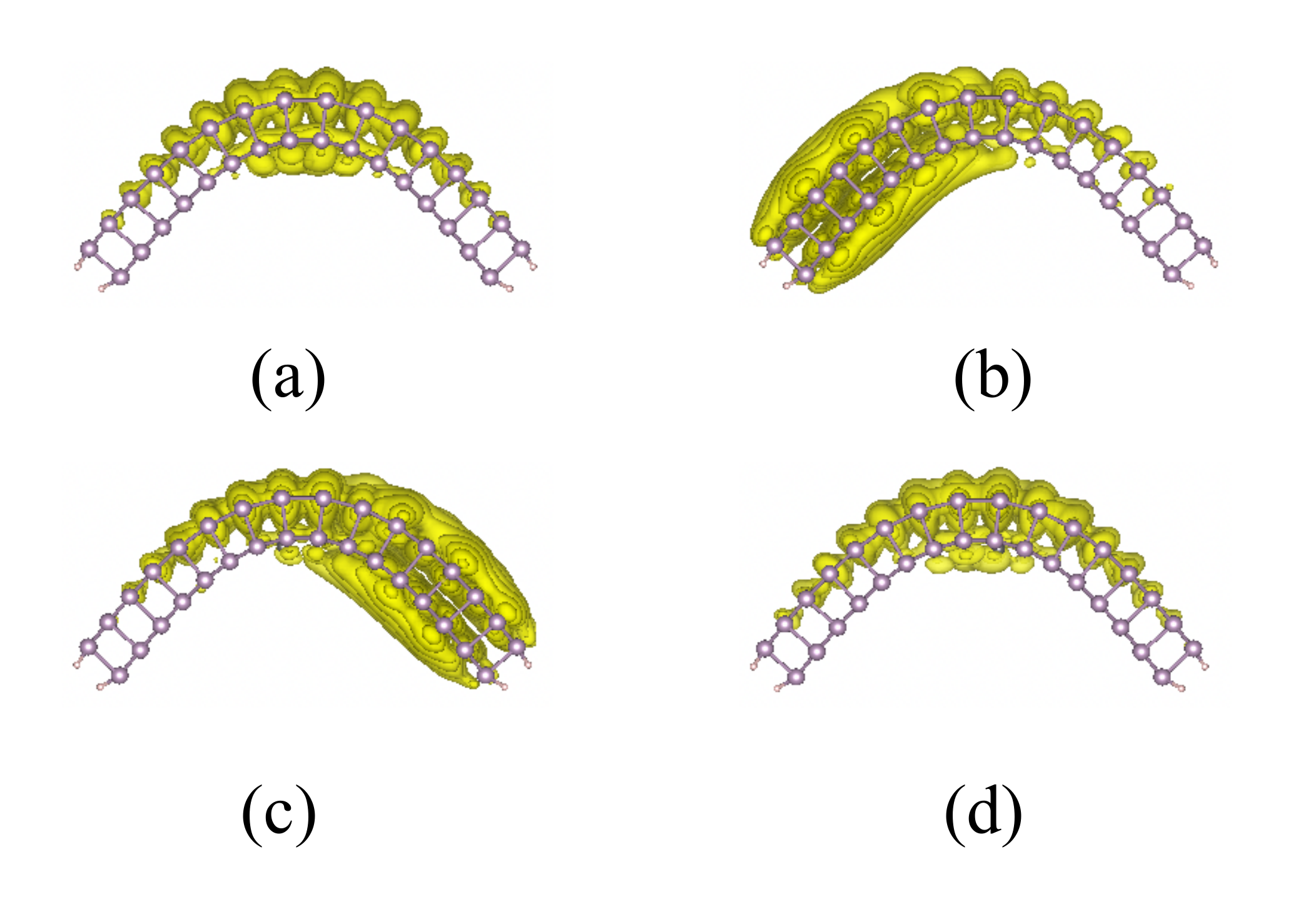}
    \caption{  The exciton wave functions at energies (a) 1.609 eV, (b) 1.615 eV, (c) 1.656, and (d) 1.662 eV are plotted for APNR at R = 10 \AA. These four bright excitons contribute to form absorption peak  A$^{\prime\prime\prime}$ .}
\end{figure}

\subsection{Phosphorene nanoribbons}

The GW-BSE calculations of APNR at R=10 \AA ~~ show that the excitons at energies 1.609 eV and 1.662 eV are formed from the transition from the valence bands to C1 (edge band) near Xi. The excitonic distribution of wave functions can be seen in Fig S5 (a) and (d), localized in the middle ribbon region, which can be understood from the partial charge analysis shown in Fig 3 (d) in the main text. The excitons at energies 1.615 eV and 1.656 eV are mainly formed from the transition from the valence bands V1, V2 to the bulk conduction bands C2, C3 near $\Gamma$. The excitonic distribution of wave functions is present in Fig S5 (b) and S5 (c), localized along the ribbon edge region, which can be understood from the partial charge analysis shown in Fig 3 (c) in the main text. Since the absorption peak A$^{\prime\prime\prime\prime}$  is formed from the combination of these four excitons, the excitonic wave functions of peak A$^{\prime\prime\prime\prime}$ are distributed throughout the ribbon width. Table S IV represents the transition probabilities with the oscillator strength near Xi and near $\Gamma$. The oscillator strength is high where the significant contribution comes from a transition near $\Gamma$ (i.e., $\sim$ 80\%). The excitons at energies 1.615 eV and 1.656 eV are formed from the combination of both transitions  from V1, V2 to C2, V3 near $\Gamma$ (i.e., $\sim$ 80\%) and   from V1 to C1, C2, C3 near Xi (i.e., $\sim$ 20\%). The main reason behind this is the quasiparticle gap at Xi (from C1 to V1 at Xi), and $\Gamma$ (from V1 to C2 at $\Gamma$), with values 2.811 eV and 2.854 eV, respectively. Those quasiparticle gaps are very close, resulting in the mixed transitions near $\Gamma$  and Xi when forming a bound exciton.

\begin{table}[h!]
 \renewcommand\thetable{S IV}
\caption {Calculated oscillator strengths $f_{i}$, exciton energies, from the transition near Xi, from the transition near $\Gamma$, and transition probabilities of the first optical absorption peak A$^{\prime\prime\prime\prime}$  of APNR at R = 10 \AA.}
\begin{tabular}{ccccc}
\hline
\hline
                          &                        & \multicolumn{2}{l}{\textbf{~~~~~Transition}} &                        \\
\hline
\textbf{Oscillator Strength ($f_{i}$)}      & \textbf{Exciton energy (eV)}            & \textbf{near Xi }     & \textbf{near $\Gamma$}      & \textbf{Transition probability} \\
\hline
\multirow{3}{*}{9.43} & \multirow{3}{*}{1.609} & V1$\rightarrow$C1      &                 & 0.26                   \\
                          &                        & V2$\rightarrow$C1      &                 & 0.52                   \\
                          &                        & V3$\rightarrow$C1      &                 & 0.17                   \\
                          \hline
\multirow{5}{*}{425.75} & \multirow{5}{*}{1.615} & V1$\rightarrow$C1      &                 & 0.09                   \\
                          &                        & V2$\rightarrow$C1      &                 & 0.07                   \\
                          &                        &              & V1$\rightarrow$C2         & 0.31                   \\
                          &                        &              & V1$\rightarrow$C3         & 0.18                   \\
                          &                        &              & V2$\rightarrow$C2         & 0.16                   \\
                          \hline
\multirow{6}{*}{671.40} & \multirow{6}{*}{1.656} & V1$\rightarrow$C1      &                 & 0.09                   \\
                          &                        & V2$\rightarrow$C1      &                 & 0.07                   \\
                          &                        & V3$\rightarrow$C1      &                 & 0.03                   \\
                          &                        &              &  V1$\rightarrow$C2       & 0.19                   \\
                          &                        &              &  V1$\rightarrow$ C3     & 0.23                   \\
                          &                        &              & V2$\rightarrow$C2         & 0.20                    \\
                          \hline
\multirow{3}{*}{40.71}   & \multirow{3}{*}{1.662} & V1$\rightarrow$C1      &                 & 0.30                   \\
                          &                        & V2$\rightarrow$C1      &                 & 0.34                   \\
                          &                        & V3$\rightarrow$C1      &                 & 0.20    \\      
\hline
\hline
\end{tabular}
\end{table}

\begin{table}[h!]
 \renewcommand\thetable{S V}
\caption { Quasiparticle  gaps $G_{0}W_{0}@PBE$ at $\Gamma$, the lowest absorption peak $E_{AI}$, and the excitonic binding energy for the lowest absoprtion peak ($E_{b}$= $G_{0}W_{0}@PBE$ - $E_{AI}$) of zigzag phosphorene at different radii of curvature (R). }
\begin{tabular}{cccc}
\hline
R (\AA)    & $G_{0}W_{0}@PBE$ (eV)   & $E_{AI}$ (eV)   &  $E_{b}$ (eV)  \\
\hline
\hline
$\infty$  & 2.65 & 1.96 & 0.69 \\
12.5 & 2.38 & 1.76 & 0.62 \\
6.67 & 1.90 & 1.46 & 0.44\\
\hline
\end{tabular}
\end{table}

\begin{table}[h!]
 \renewcommand\thetable{S VI}
\caption {The energy absorption peaks for spin singlet ( E$_{A}^{S}$), spin triplet (E$_{A}^{T}$), and spin singlet-triplet splitting ($\Delta$$^{S-T}$ = E$_{A}^{S}$- E$_{A}^{T}$)  at different radii of curvature (R) for APNR.}
\begin{tabular}{ccccc}
\hline
\hline
\textbf{R} (\AA)   & \textbf{Peaks} & \textbf{E$_{A}^{S}$ (eV)} & \textbf{E$_{A}^{T}$ (eV)} & \textbf{$\Delta$$^{S-T}$ (eV)} \\
\hline
             & A              & 1.31                & 1.27                & 0.04                                     \\
Flat & B              & 1.66                & 1.66                & 0.00                                        \\
             & C              & 1.85                & 1.82                & 0.03                                     \\
\hline
  
          & A$^{\prime}$              & 1.26                & 1.22                & 0.04                                     \\
  15           & B$^{\prime}$             & 1.56                & 1.55                & 0.01                                     \\
             & C $^{\prime}$             & 1.72                & 1.70                 & 0.02                                     \\
\hline
         & A$^{\prime\prime}$            & 1.55                & 1.50                 & 0.05                                     \\
12.5           & B$^{\prime\prime}$           & 1.77                & 1.74                & 0.03                                     \\
             & C$^{\prime\prime}$            & 2.48                & 2.43                & 0.05                
             \\
\hline
           & A$^{\prime\prime\prime}$            & 1.67                & 1.57                & 0.10 \\

 11            & B$^{\prime\prime\prime}$           & 2.03                & 1.96                & 0.07                                     \\
             & C$^{\prime\prime\prime}$            & 2.45                & 2.44                & 0.01                                     \\
\hline
           & A$^{\prime\prime\prime\prime}$          & 1.65                & 1.60                 & 0.05                                     \\
   10          & B$^{\prime\prime\prime\prime}$         & 2.06                & 2.01                & 0.05                                     \\
             & C$^{\prime\prime\prime\prime}$          & 2.51                & 2.49                & 0.02                                     \\
\hline
\hline
\end{tabular}
\end{table}

\begin{table}[h!]
 \renewcommand\thetable{S VII}
\caption {The energy absorption peaks for spin singlet ( E$_{A}^{S}$), spin triplet (E$_{A}^{T}$), and spin singlet-triplet splitting ($\Delta$$^{S-T}$ = E$_{A}^{S}$- E$_{A}^{T}$)  at different radii of curvature (R) for ZPNR.}
\begin{tabular}{ccccc}
\hline
\hline
\textbf{R} (\AA)   & \textbf{Peaks} & \textbf{E$_{A}^{S}$ (eV)} & \textbf{E$_{A}^{T}$ (eV)} & \textbf{$\Delta$$^{S-T}$ (eV)} \\
\hline
             & A              & 1.96                & 1.95                & 0.01                                     \\
Flat & B              & 2.17                & 2.17                & 0.00                                        \\
             & C              & 2.55                & 2.55                & 0.00                                    \\
\hline
  
          & A$^{\prime}$             & 1.76                & 1.76               & 0.00                                     \\
  12.5           & B$^{\prime}$             & 1.93                & 1.93                & 0.00                                     \\
             & C$^{\prime}$             & 2.18                & 2.18                 & 0.00                                     \\
\hline
         & A$^{\prime\prime}$            & 1.46                & 1.45                 & 0.01                                     \\
6.67           & B$^{\prime\prime}$            & 1.71                & 1.69               & 0.02                                     \\
 
\hline
\hline
\end{tabular}
\end{table}

\begin{figure}[h!]
  \renewcommand\thefigure{S6}
    \includegraphics[scale=0.45]{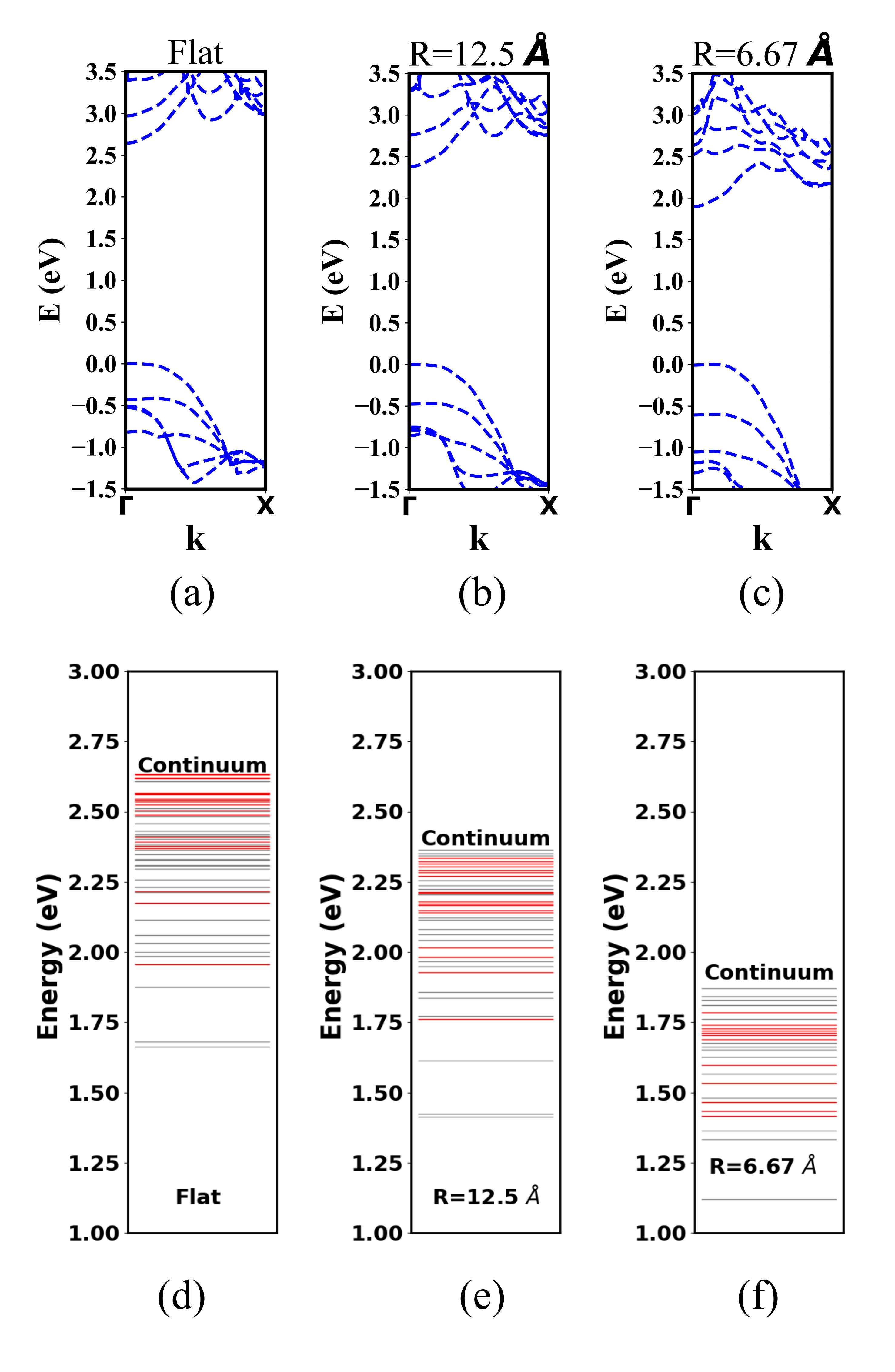}
    \caption{(Upper panel) G$_0$W$_0$ band structures of ZPNR at different radii of curvature. (Lower panel) excitonic energy levels of armchair ZPNR at different radii of curvature. }
\end{figure}

\end{document}